\documentclass[11pt]{article}
\def\lesssim{\mathrel{\hbox{\rlap{\hbox{\lower4pt\hbox{$\sim$}}}\hbox{$<$}}}}
\let\la=\lesssim
\def\gtrsim{\mathrel{\hbox{\rlap{\hbox{\lower4pt\hbox{$\sim$}}}\hbox{$>$}}}}
\let\ga=\gtrsim

\begin{document}

\title{Old models for Cygnus X-1 and AGN}
\author{G. S. Bisnovatyi-Kogan and S. I. Blinnikov}
\date{}

\begin{titlepage}

\maketitle
\thispagestyle{empty}
\begin{abstract}
Recently, there appeared many papers devoted to the modeling of X-ray properties of
Cygnus X-1 and other  black hole accretion disk candidates: e.g.
J.Poutanen, J.Krolik, F.Ryde, MNRAS 292 (1997) L21;
E.Agol, J.Krolik, ApJ 507 (1998) 304;
A.Beloborodov, in ``High Energy Processes in Accreting Black Holes'',
ASP Conf. Series, 161, (1999), p.295. 
The goal of this electronic publication is to draw attention to our old papers where many
ideas of recent discussions were anticipated (hot coronae, Comptonization,
photon damping of waves, particle acceleration and matter ejection from accretion disks
with large scale poloidal magnetic fields, etc.)
\end{abstract}
\end{titlepage}
\begin{center}

{\Large {\bf A hot corona around a black-hole accretion
disk as a model for Cygnus X-1}
\bigskip

G. S. Bisnovatyi-Kogan and S. I. Blinnikov}

\bigskip
{\em Institute for Space Research, USSR Academy of Sciences, Moscow}

(Submitted April 15, 1976)

Pis'ma Astron. Zh. 2, 489-493 (October 1976)

Sov.Astron.Lett. 2, 191-193  (Sep.-Oct. 1976)

\end{center}

\begin{small}
Heat transfer in the region of maximum energy release
of an accretion disk will take place mainly by
convection, serving to enhance the turbulence and to generate
a powerful acoustic flux. The hard X-rays
emitted by Cyg X-1 ($E \la 200$ keV) might result from
Comptonization of soft photons in a corona formed
around the disk through this heating.

\end{small}

\bigskip

 With high probability, the X-rays emitted by the source
Cygnus X-1 come from a black hole that is in a regime
of disk accretion. The theory of disk accretion has been
developed by several authors [1]--[4]. But difficulties arise if
one seeks to explain the spectrum of Cyg X-1 on the basis
of this theory. The source has a luminosity $L \la 10^{38}$
erg/sec $\approx 0.1 L_c$, where $L_c = 1.3 \cdot 10^{38} M/M_{\odot}$
is the critical Eddington luminosity. For this luminosity the
theoretical spectrum should fall off sharply in the energy range
$h\nu \approx 7$ keV, whereas the observed spectrum extends up to
energies as high [5,6], as $ \approx 200$ keV, or even beyond [7,8].

To overcome this difficulty, models have been proposed [9,10]
wherein the properties of the radiating regions
are essentially determined by the observations. An attempt has been
made [11] to construct a self-consistent model
for Cyg X-1 that would account for the hard part of the
radiation. This model, which neglects radiation pressure
from the very outset, appears to have an internal contradiction in
that it implies a disk as thick as its radius and drift velocity
comparable to the orbital velocity.

Hard radiation up to 200-keV energy can be present
in the spectrum of an accretion disk only in the event that
regions with hot $(T_e \approx 10^9$ K) electrons participate
in forming the spectrum; harder photons would require still
more energetic particles. In an analysis of accretion disks
we have shown [12] that hot regions or coronae with
$T_e \approx 10^9$ K will form around them.

The standard theory presupposes that radiative heat
conduction is responsible for energy transfer in the vertical
direction. In the region of maximum energy release,
$P \approx P_r \gg P_g$, $\kappa = \kappa_{\rm es}$, and the vertical
structure can easily be determined analytically. The density in this
region is found [3] to be independent of $z$, while the
temperature declines toward higher $z$. Such a situation will
evidently be convectively unstable, since the entropy will
decrease in the vertical direction. The onset of convective
instability will serve to equalize the entropy. The mean
density in the disk will become approximately an order of
magnitude higher than the density of the disk in the
standard theory.

The convective heat flow $Q_{\rm conv}$, the convection velocity $v$,
and the excess $\Delta\nabla T$ of the temperature
gradient above  the adiabatic gradient (we shall take the mixing
length $l$ to be the half-thickness $z_0$ of the disk) are given
[13] by the equations

\begin{equation}
Q_{\rm conv} = C_p \rho v (z_0 /2) \Delta\nabla T ,
\end{equation}

\begin{equation}
v = \left[ \frac{Q_{\rm conv}(1 + 4P_{r}/P_g)}{2\rho C_p T}
\frac{GM}{R^3}\right]^{1/3} z_0^{2/3},
\end{equation}

\begin{equation}
\Delta\nabla T = \left(\frac{4Q_{\rm conv}}{\rho C_p}  \right)^{2/3}
\left[ \frac{T}{(1 + 4P_r/P-g)GM}\right]^{1/3}
 R z_0^{-5/3}.
\end{equation}

Hence we readily find that the excess
$\Delta\nabla T$ in the temperature gradient
amounts to no more than $20\%$ of  $\nabla T$, while
the convection velocity is $v = 3 \cdot 10^8$ cm/sec for $r = 10 r_g$,
a value close to the velocity $v_s$ of sound in this region.
The heat flow is carried mainly by convection $(Q \approx Q_{\rm conv})$;
because of the high density the radiative flux is equal to
$\approx 20\%$ of the total flux.

In the convective disk the flow $Q_{\rm ac}$ of acoustic energy
in the vertical direction is given [14] by
\begin{equation}
 Q_{\rm ac} \simeq \rho v^3 (v/v_s)^5 \simeq 10^{21-22}
 \mbox{erg}\cdot \mbox{sec}^{-1} \cdot \mbox{cm}^2
\end{equation}
and is a quantity of the same order as the total energy
flux. An amount of order $(P_g/P_r)Q_{\rm ac}$ is expended
in heating the outer layers to an optical depth $\tau < 1$.
One other mechanism serves to heat these layers [12].
A particle in a region near the surface of the disk that is transparent to radiation will be subject to the influence of radiation from
the whole disk, and not only to the local radiation pressure gradient. The force of the radiation pressures will
accelerate particles in the transparent region above the
disk. Calculations of the equations of motion for particles
subject to radiative, centrifugal, and gravitational forces
show that in the region with $\tau < 1$ the particles (protons
and electrons) will acquire vertical velocities corresponding to the temperatures
$(1-8)\cdot 10^8$ K of protons for
$L \approx 0.1 L_c$.
Turbulent relaxation of the particle motions in
the corona will tend to equalize the mean energies of the
ions and electrons and to form a quasi-Maxwellian
particle velocity distribution. Thus the combined action of the
two heating mechanisms we have described will produce
around the accretion disk a hot corona with $T_e \approx 10^9$ K
for $L \approx 0.1L_c$. The existence of an analogous corona has
been postulated phenomenologically by Price and Liang [10].

We now estimate the density $\rho_{\rm co}$ of the corona and
the amount of material it contains. Since the gas
 pressure varies continuously with transition from the
photosphere to the corona, we readily find that at the base of
the corona

\begin{equation}
\rho_{\rm co} \simeq \rho_s T_s/T_{\rm co} \simeq 10^{-2} \rho_s
\approx 10^{-5} \mbox{g/cm}^3,
\end{equation}
where $\rho_s$, $T_s$ denote the density and temperature in the
photosphere of the disk for the parameters of the source
Cyg X-1 in the region of maximum energy release. The
surface density of the corona is determined by the
condition $\tau _{\rm es} \approx 1$ and is $\approx 2 \, \mbox{g/cm}^2$,
which is more than an order of magnitude lower than the density of the opaque
disk.

The hot corona readily affords an explanation of the
peculiarities of the Cyg X-1 spectrum. The soft X-rays
at $h\nu \la 7$ keV, which comprise $\approx 70\%$ of the total flux,
are formed in the photosphere of the opaque disk. Some of
the radiation $(\approx 10\%)$ will pass through the hot corona,
 and Comptonization will generate hard radiation up to
$h\nu \approx 3kT_e \approx 200$ keV, which will amount
to $\approx 30\%$ of the total flux. In Cyg X-1 the luminosity varies around $L = 0.1 L_c$. Under these conditions the region of the disk with
$P_r \gg P_g$, where convective heating is important, will be
spatially small and will convert $\approx 10\%$ of the soft
radiation flux, in accord with the requirements that the
observations impose on the model.

The changes in the spectrum as the luminosity of
Cyg X-1 varies [6,15,16] exhibit the following characteristic
behavior. As the total energy flux rises, the radiation in
the soft part of the spectrum ($h\nu \la 7$ keV) increases, but
in the hard range ($h\nu \ga 10$ keV) it remains almost
constant, or perhaps may even decrease slightly. We shall
assume that the variations in the luminosity are associated
with fluctuations in the power of accretion. As the mass
flow $\dot M$ rises in the region with $P_r \gg P_g$ the fraction of
the acoustic flow (4) expended in heating the corona will
decrease:

\begin{equation}
(P_g/P_r)Q_{\rm ac} \sim \dot M^{-1}.
\end{equation}

For $L \approx 0.1L_c$, when acoustic heating predominates,
the rise in $\dot M$ may cause some decrease in the heating
of the corona and in the amount of hard radiation. At the
same time the flux in the soft range is determined by the
radiation of the disk photosphere and will increase with
$\dot M$. It is worth noting that in strong bursts of luminosity,
when $L$ reaches about $0.3L_c$, the heating will begin to
be governed by radiation-pressure forces, so the
temperature of the corona and thereby also the power of
the hard radiation should increase along with the rise in $\dot M$
and in the total energy flux.

The observational evidence for the spectrum at
$h\nu > 150$ keV is considerably less reliable than for the soft
range. For instance, Baker et al. [8] have measured the
radiation to $h\nu \approx 10$ MeV, but find an observed signal of
only $\approx 1\%$ of the background (they also report a deficiency
of $\gamma$-ray photons in the direction of Cyg X-1). In the range
up to 600 keV, Haymes and Harnden [7]  find a break in the
spectrum at $h\nu  \approx 150$ keV, and the spectral index $\alpha$ changes from $\alpha = 1.9$ at $h\nu < 124$ keV to
$\alpha = 3.1$ at $h\nu > 154$ keV.

Even if a corona with $T \simeq 10^9$ K is present in the disk
accretion model, radiation cannot be formed at
$h\nu = 200-2000$ keV. Electrons with $T_e = 10^{10}$ K or
fast nonthermal electrons would be needed for that purpose.
There are in fact two ways to form such fast electrons.
Both would require the presence of a magnetic field in the disk.
A magnetic field could exist in the disk either through
twisting of the lines of force by differential rotation [3,17,18],
or through infall onto the black hole of magnetized
material having a small angular momentum [19]. In the latter
case a poloidal magnetic field would be generated.

In the binary system containing the source Cyg X-1,
some of the material flowing from the giant star is
dispersed in space near the system. The attraction of the
black hole will not only produce an accretion disk.
A small proportion of material having a low angular momentum
will fall into the black hole and be decelerated in the
disk. If this deceleration takes place at radii of
$(10-30)r_g$ and if a thin collisionless shock wave is formed
(as is very likely in the presence of an azimuthal magnetic field)
wherein the kinetic energy is transformed into thermal
energy and $T_e \approx T_i$, then hot electrons with
$T= 10^{10} - 10^{12}$ K
will appear. The inverse Compton mechanism of interaction
of the disk radiation with these electrons can
lead to the generation of hard radiation with $h\nu \approx
200-2000$ keV or even higher energies.

Another mechanism for producing fast particles is
analogous to the pulsar process. If magnetized matter
with low angular momentum falls into the black hole (in
addition to the disk accretion), a strong poloidal
magnetic field will arise [19]. By analogy to pulsars [20],
rotation will generate an electric field of strength
$E \approx -(v/c)B$ in
which electrons are accelerated to energies
$\varepsilon \approx R(v/c)Be \approx 3\cdot 10^4 [B/(10^7\,\mbox{Gauss})]$ Mev
where $v/c \approx 0.1$ and $R \approx 10^7$ cm
 is the characteristic scale. In a field $B \approx 10^7$ Gauss,
such electrons will generate synchrotron radiation with
energies up to $\approx 10^5$ keV. Just as in pulsars, it would be
possible here for $e^{+}e^{-}$  pairs to be formed and to
participate in the synchrotron radiation.

The authors express their appreciation to
Ya.B. Zel'dovich, A. F. Illarionov, and I. S. Shklovskii for
helpful discussions and valuable comments.

\bigskip
\hrule
\bigskip

1. N. I. Shakura, ``A disk model for accretion of gas by a relativistic
star in a close binary system",
Astron. Zh. {\bf 49}, 921-929 (1972) [Sov. Astron.{\bf 16},756-762 (1973)].

2. J. E. Pringle and M.J.Rees, ``Accretion disk models
for compact X-ray sources", Astron. Astrophys. {\bf 21},
1-9 (1972).

3. N. I. Shakura and R. A. Syunyaev, ``Black holes in binary systems: observational appearance", Astron. Astrophys. {\bf 24}, 337-355 (1973).

4. I. D. Novikov and K. S. Thorne, ``Black-hole astrophysics",
 in: Black Holes (Les Houches lectures, Aug. 1972), Gordon and Breach (1973), pp.343-450.

5. F. Frontera and F. Fuligni, ``Energy-spectrum variability
 of Cyg X-1 in hard X-rays", Astrophys. J. {\bf 196},
597-599 (1975).

6. G. F. Carpenter, M. J. Coe, A. R. Engel, and J. J. Quenby, ``Ariel 5 hard X-ray measurements of galactic and extragalactic source spectra", Proc.14th Intl. Cosmic Ray Conf. (Munich) {\bf 1}, 174-179 (1975).

7. R. C. Haymes and F. L. Harnden, ``Low-energy $y$ radiation
from Cygnus" Astrophys. J. {\bf 159}, 1111-1114 (1970).

8. K. E. Baker, R. L. Lovett, K. J. Orford, and D. Ramsden,
``Gamma rays of 1-10 MeV from the Crab and Cygnus regions",
Nature Phys. Sci. {\bf 245}, 18-19 (1973).

9. K. S. Thorne and R. H. Price,
``Cyg X-1: an interpretation of the
spectrum and its variability", Astrophys. J. {\bf 195},
 L101-L105 (1975).

10. R. H. Price and E. P. Liang, Preprint (1975).

\noindent
[published in Astrophys. J., {\bf 218}, Nov. 15, 1977, 247-252.]

11. S. L. Shapiro, A. P. Lightman, and D. M. Eardley,
``A two-temperature accretion-disk model for Cyg X-l",
Astrophys. J. {\bf 204}, 187-199 (1976).

12. G. S. Bisnovatyi-Kogan and S. I. Blinnikov,
Preprint Inst. Kosmich. Issled.
Akad. Nauk SSSR No. 271 (1976); Astron. Astrophys. (in press).

\noindent
[published in A\&A 1977, {\bf 59}, 111-125]

13. M. Schwarzschild,
Structure and Evolution of the Stars, Princeton Univ.
Press. (1958).

14. L. Biermann and R. Lust. ``Nonthermal phenomena in
stellar atmospheres",
in: Stellar Atmospheres, Univ. Chicago Press (1960), Chap. 6.

15. D. R. Parsignault, A. Epstein, J. E. Grindlay,
 E. J. Schreier, H. Schnopper,
H. Gursky, Y. Tanaka, A. C. Brinkman, J. Heise, 1. Schrijver, R. Mewe,
E. Gronenschild, and A. den Boggende, ``ANS observations of Cyg X-1",
Astrophys. Space Sci. {\bf 42}, 175-184 (1976).

16. S. S. Holt, E. A. Boldt, L. J. Kaluzienski, and P. J. Serlemitsos,
``Observations of a new transition in the emission
from Cyg X-1", Nature {\bf 256}, 108-109 (1975).

17. L. A. Pustil'nik and V. F. Shvartsman,
``Possible influence of magnetic fields on the structure of a plasma
accretion disk in binary systems", in:
Gravitational Radiation and Gravitational Collapse (IAU Sympos. No. 64),
Reidel (1974), p. 213.

18. D. M. Eardley and A. P. Lightman,
``Magnetic viscosity in relativistic
 accretion disks", Astrophys. J. {\bf 200}, 187-203 (1975).

19. G. S. Bisnovatyi-Kogan and A. A. Ruzmaikin,
``The accretion of matter
by a collapsing star in the presence of a magnetic field",
Astrophys. Space Sci. {\bf 42}, 401-424 (1976).

20. P. Goldreich and W. H. Julian, ``Pulsar electrodynamics",
Astrophys. J., {\bf 157}, 869-880 (1969).

\newpage
\vglue 1cm
\begin{center}
{\Large {\bf Models for the X-ray brightness fluctuations in
Cygnus X-1 and active galaxy nuclei}
\bigskip

G. S. Bisnovatyi-Kogan and S. I. Blinnikov}
\bigskip

{\em Institute for Space Research, USSR Academy of Sciences, Moscow}

(Submitted April 7, 1978)

Pis'ma Astron. Zh. 4, 540-543 (December 1978)

Sov.Astron.Lett. 4, 290-291 (Nov.-Dec. 1978)
\end{center}

\begin{small}
The X-ray brightness fluctuations induced in
Cyg X-l and active galaxy nuclei by
convection and turbulence in the
sub-photospheric layers are discussed in terms of
the disk-accretion and supermassive-star
models. The variability time scales should be
comparable with those observed, but in the case of
supermassive stars and disks it is difficult to obtain the
observed amplitude of the fluctuations.

\end{small}

\bigskip

1. One noteworthy property of the X-ray source Cygnus X-l is the
variability of its flux on time scales ranging from milliseconds to a
fraction of a second [1].
This variability is generally attributed to the
presence of a turbulent accretion disk in the Cyg X-l system. In
particular, the flux variations have been explained by the rotation of a
hot spot [2,1] or alternatively, by the onset of instability in a zone where
radiation pressure predominates [3,5].

In this letter we shall consider another model for the variability.
We shall outline more fully what properties of the flux variations are to
be expected on the accretion disk model as a direct result of the
presence of turbulence, and how convective instability should develop
in a region of high radiation pressure [6,7]. Acoustic waves generated in
the convection zone will escape into optically thin layers, and will not
only induce variable soft X-rays in the photosphere, but will also be
responsible for variable heating of the corona. Comptonization of the
photospheric radiation by the hot electrons under conditions where both
temperature and density are variable will lead to flux variations in the
hard range, ${h\nu} \ga $ 5 keV. This X-ray fluctuation mechanism, involving
the emergence of waves into transparent layers, evidently is of the same
wave nature as the variability of the ultraviolet excess in stars
experiencing intensive convection, such as T Tauri and UV Ceti.

2. The waves escaping into the transparent layers and producing
variable radiation in the photosphere and corona will occupy a rather
narrow frequency band. Physically, the reason for this circumstance is
that media with a high radiation pressure and a nonuniform distribution
of plasma along the $z$ coordinate (across the disk) will serve as efficient
filters, isolating a characteristic frequency range from the broader
spectrum generated by convection and turbulence. Waves of low
frequency and a wavelength exceeding the scale height of the
atmosphere will not escape outside but will induce oscillations of
the coronal atmosphere as a whole. On the other hand, under conditions
where radiation pressure predominates, high-frequency waves will
experience very severe damping because of radiative friction, and their
role in heating the corona will be insignificant. High frequencies may,
however, appear in the observations either through nonlinear
transformation of low-frequency waves or due to inadequate processing of
the observational data [8].

We have examined elsewhere [9] the propagation of waves through
a medium with strong radiation pressure, followed by their escape into
the atmosphere. Waves emerging into the transparent layers will have a
phase and group velocity equal to the velocity of sound in gas,
$v_g = (\gamma P_g /\rho)^{1/2} = (\gamma{\cal R}T)^{1/2}$,
where $\gamma = $5/3 or $\gamma =$ 1 according as scattering
or absorption predominates, ${\cal R}$ is the gas constant,
and the temperature  $T$  of the equilibrium atmosphere is
approximately equal to the temperature $T_{\rm e}$ of the photosphere.
The characteristic frequency $\omega_{\rm c}$ of waves
emerging into the atmosphere is given by the expression
\begin{equation}
\omega_{\rm c}= \left(\frac{\gamma}{{\cal R}T}\right)^{1/2}
 g \left(1 - \frac{H}{H_c}\right) {\rm sec}^{-1} .
\end{equation}
For the accretion disk model, the gravitational acceleration $g$ at radius
$r$ is equal [10] to
\begin{equation}
 g = \frac{GM}{r^2} \frac{h_0}{r} ,
\end{equation}
where the characteristic thickness is
\begin{equation}
 h_0 = 3\frac{L}{L_c} R_0\left[1-\left (\frac{R_0}{r}\right)^{1/2}\right] ,
\end{equation}
and $R_0$ is the radius of the inner edge of the disk. The quantity
$H/H_{\rm c} = \kappa_0 H/gc$  represents the ratio of the
 radiation pressure force to the
gravity; $\kappa_0$ is the opacity, including both absorption and
scattering [9].
In a spherically symmetric star of luminosity $L$ and radius $R$ we will
have
$g = GM/R^2$, $H = L/4 \pi R^2$, and $H/H_c = L/L_c$,
where $L_c = 4 \pi cGM/\kappa_0$
is the Eddington limiting luminosity.

3. To calculate the characteristic frequencies of the fluctuations
for Cyg X-l, we shall adopt the convective accretion-disk models given
in a previous paper [7] and by Shakura et al. [11].
For a black hole of mass $M = 10 M_{\odot}$, we find that in
both models as the luminosity rises from $0.1 L_c$ to $0.3 L_c$
there is little change in the characteristic frequency $\omega_c$
corresponding to the equilibrium temperature, as given by Eq. (1); the
values of $\omega_{\rm c}$ are confined [9] to the range 5-40 msec.
Note that if a corona
with $T_c \approx 10^2 T_e$ is present,
waves whose frequency is $\omega = \omega_{\rm c}$ or even
somewhat lower will be able to escape.
The frequency range mentioned
is in good accord with the observed time-scales of variability.
Our mechanism can yield fluctuations weakly
correlated in time, and can simulate the white noise derived from
observational analysis [8] of the brightness fluctuations of Cyg X-l.

Despite the good agreement between the theoretical time
characteristics and the observations, analysis of small oscillations in
our model still does not suffice to explain the amplitudes of the
variability we observe, because pulsations should, in general, occur
independently in regions whose size is comparable with the velocity of
sound multiplied by the pulsation period, or about one-tenth the
diameter of the zone of maximum energy production. Small pulsations
should accordingly be smoothed out. However, the mechanism we are
proposing could operate in a highly nonlinear regime. Strong
nonlinearity of the waves would be expected in the light of theoretical
estimates [7], and is essential to the very existence of the corona. The
frequencies obtained from linear analysis correspond to the
characteristic growth times of strong nonlinear flares. This problem
awaits further theoretical treatment. Indeed, we would point out that
not even the observational situation is fully clear [8].

An evaluation of the proportion $\eta$ of the acoustic flux that
emerges into the corona and produces heating shows [9] that as the
luminosity rises by a factor of 3 from 0.1 $L_c$ to 0.3 $L_c$,
 the value of $\eta$
will drop by a factor of 5-10. Thus, in accord with our previous
suggestion [6,7], as the luminosity increases the power of the corona will
remain approximately the same, or may even decline somewhat. This
situation will prevent the hard X-ray flux
($h\nu > 5$ keV) of Cyg X-l from
changing appreciably as the total luminosity varies, in agreement with
the observations [12].

4. Multicolor photometry of the nuclei of certain galaxies has
revealed the presence of comparatively short-period brightness
fluctuations in the nucleus of the Seyfert galaxy [13,14] NGC 4151
($\approx$ 130 days) and in the BL Lacertae object [15]
 OJ 287 ($\approx$ 180 days). There is
presently no consensus as to the nature of active galaxy and quasar
nuclei. Three models have been explored [16,17]:
a) a dense star cluster; b) a supermassive star; c) disk accretion onto a
massive black hole. The variability mechanism proposed here can
operate in the two models b and c, for in both cases the
subphotospheric layers will be strongly convective, resulting in the
formation of a corona with variable heating. Using Eq. (1) to estimate
the characteristic period of the fluctuations along with the standard
model for a supermassive star [18] we find that for a mass
$M = 10^8 M_\odot$ and a radius $100 R_{\rm g}$
($R_g = 2GM/c^2$ is the gravitational radius), the period
corresponding to $\omega_c$
would be $\approx 160$ days, a value consistent with observation. On the
accretion disk model, such a period would prevail in the zone of
maximum energy production [9] for a black hole of $ M = 10^8  M_\odot$,
if the luminosity $L = 0.1 L_c$. In galaxy nuclei the fast
component typically exhibits an
approximately constant period in conjunction with sharp changes in
phase [15], a behavior which, it would seem, accords with a convective
wave origin for such fluctuations.

It is worth emphasizing, however, that both in the supermassive
star model and in the model of an accretion disk around a
supermassive black hole, the ratio of
the radiation pressure to the gas pressure is far higher than in the case
of accretion onto a black hole of stellar mass. Acoustic waves will
therefore be damped much more strongly. Our calculations indicate [9]
that in this event the emergent acoustic flux will comprise no more
than 1$\%$ of the flux generated at large optical depth - well below the
variability amplitude observed. The supermassive star model would be
supported if strictly periodic brightness oscillations of constant phase
were detected, associated with the rotation or with pulsations of such a
star as a whole, as might be the case [14] for NGC 4151. But analysis of
observations of several variable nuclei does not reveal any strict
periodicities [19]. Most likely the observed variability should be modeled
by a random process [20] (see, however, Ozernoi et al. [21]).
 The model of a
dense star cluster with frequent supernova outbursts offers the best fit,
in our opinion, to the irregular variability of galaxy nuclei.

\bigskip
\hrule
\bigskip

1. E. Boldt, ``X-ray signatures: new time scales and spectral features," in:
Eighth Texas Sympos. on Relativistic Astrophysics. Ann. New York Acad.
Sci. {\bf 302}, 329-348 (1977).

2. R. A. Syunyaev, ``Variability of X-rays from black holes with accretion disks,"
Astron.Zh. {\bf 49}, 1153-57 (1972) [Sov. Astron. {\bf 16}, 941-944 (1973)].

3. A. P. Lightman and D. M. Eardley, ``Black holes in binary systems:
instability of disk accretion," Astrophys. J. {\bf 187}, L1-L3 (1974).

4. N. Shibazaki and R. Hoshi. ``Structure and stability of an accretion disk around a black
hole," Progr. Theor. Phys. {\bf 54}, 706-718 (1975).

5. N. I. Shakura and R. A. Syunyaev, ``A theory of the instability of disk
accretion onto black holes," Mon. Not. R. Astron. Soc. {\bf 175}, 613-632 (1976).

6. G. S. Bisnovatyi-Kogan and S. I. Blinnikov, ``A hot corona around a black-hole accretion
disk as a model for Cyg X-l," Pis'ma Astron. Zh. {\bf 2}, 489-493 (1976) [Sov. Astron Lett.
{\bf 2}, 191-193 (1977)].

7. G. S. Bisnovatyi-Kogan and S. I. Blinnikov, ``Disk accretion onto a black
hole at subcritical luminosity," Astron. Astrophys. {\bf 59}, 111-125 (1977).

8. M. C. Weisskopf and P. G. Sutherland, ``On the physical reality of the
millisecond bursts in Cyg X-l," Astrophys. J. {\bf 221}, 228-233 (1978).

9. G. S. Bisnovatyi-Kogan and S. I. Blinnikov, ``Wave propagation in a
medium with high radiation pressure" [in Russian]. Preprint Inst. Kosmich.
Issled. Akad. Nauk SSSR No. 421 (1978).

\noindent
[Published in Astrophysics  {\bf 14}, 316-325 (1978); {\bf 15}, 99-107 (1979).]

10. N. I. Shakura and R. A. Syunyaev. ``Black
holes in binary systems: observational appearance," Astron. Astrophys. {\bf 24},
337-355 (1973).

11. N. I. Shakura, R. A. Syunyaev, and S. S. Zilitinkevicn, ``Turbulent
energy transport in accretion disks," Astron. Astrophys. {\bf 62}, 179-187 (1978).

12. J. F. Dolan, C. J. Crannell, B. R. Dennis, K. J. Frost, and L. E. Orwig,
``Intensity transitions in Cyg X-l observed at high energies from OSO-8,"
Nature {\bf 267}, 813-815 (1977).

13. V. M. Lyutyi and A. M. Cherepashchuk, ``H$\alpha$ emission variability in
the Seyfert nuclei NGC 4151, 3516, 1068" [in Russian], Astron. Tsirk.
No. 831. 1-3 (1974).

14. E. T. Belokon', M. K. Babadzhanyants, and V. M. Lyutyi,
``Periodicity of the Seyfert galaxy NGC 4151 in the optical," Astron. Astrophys. Suppl.
{\bf 31}, 383 (1978).

l5. V. M. Lyutyi, ``Optical variability of OJ 287" [in Russian]. Peremennye
Zvezdy {\bf 20}, 243-249 (1976).

16. V. L. Ginzburg and L. M. Ozernoi, ``The nature of quasars
and active galaxy nuclei," Astrophys. Space Sci. {\bf 50}, 23-41 (1977).

17. M. J. Rees, ``Quasar theories,"
in: Eighth Texas Sympos. on Relativistic
Astrophysics. Ann. New York Acad. Sci. {\bf 302}, 613-636 (1977).

18. Ya. B. Zel'dovich and I. D. Novikov, Relativistic Astrophysics, Vol. 1,
Stars and Relativity, Univ. Chicago Press (1971).

19. M. M. Basko and V. M. Lyutyi,
``Search for periodic optical variations of the Seyfert nuclei NGC 1275 and 3516,"
Pis'ma Astron.Zh. {\bf 3}, 104-103 (1977) [Sov. Astron. Lett. {\bf 3}, 54-56 (1977)].

20. W. H. Press, ``Flicker noises in astronomy and elsewhere," Comments
Astrophys. Space Phys. {\bf 7}, 103-119 (1978).

21. L. M. Ozernoi, V. E. Chertoprud, and L. I. Gudzenko, ``Comments on
the light curve of the quasar 3C 273," Astrophys. J. {\bf 216}, 237-243 (1977).

\end{document}